%% file: acl_latex.tex
\documentclass[11pt]{article}

\usepackage[final]{acl}

\usepackage{times}
\usepackage{tcolorbox}
\usepackage{latexsym}
\usepackage{booktabs}   
\usepackage{multirow}   
\usepackage{amsmath}
\usepackage{amssymb}    
\usepackage{bm}         
\usepackage{enumitem}
\usepackage{url}       
\usepackage{hyperref}  
\usepackage{xurl} 
\usepackage{caption} 
\usepackage{todonotes}
\usepackage[table]{xcolor}
\usepackage[T1]{fontenc}

\usepackage[utf8]{inputenc}

\usepackage{microtype}

\usepackage{inconsolata}

\usepackage{graphicx}

\title{RankSteer: Can Pointwise LLM Rankers Be Calibrated at the Representation Level?}

\author{
  Yumeng Wang \\
  Leiden University \\
  Leiden, The Netherlands \\
  {\small \texttt{y.wang@liacs.leidenuniv.nl}}
  \And
  Catherine Chen \\
  Brown University \\
  Providence, RI, USA \\
  {\small \texttt{catherine\_s\_chen@brown.edu}}
  \And
  Suzan Verberne \\
  Leiden University \\
  Leiden, The Netherlands \\
  {\small \texttt{s.verberne@liacs.leidenuniv.nl}}
}

\begin{document}
\maketitle
\begin{abstract}
Large language models (LLMs) are strong zero-shot pointwise rankers, but lag behind pairwise and listwise methods. Beyond missing comparative signals, we identify a \textit{calibration gap}: ranking-relevant information encoded in hidden states is not fully captured by the scalar output head.
We propose RankSteer, a post-hoc activation-steering framework that calibrates ranking via projection-based interventions along multiple directions at inference time: decision, evidence, and, optionally, role. This is achieved without updating model weights or introducing cross-document comparisons.
We instantiate RankSteer on two structurally distinct pointwise variants 
and observe improvements over their respective baselines on most TREC DL and BEIR datasets across three backbones. This suggests that the calibration gap is a general property of pointwise rankers.
Our additional geometric analysis shows that steering improves ranking by concentrating each query's document representations along an existing ranking geometry, offering new insight into how LLMs internally represent and calibrate relevance judgments.

\end{abstract}

\section{Introduction}
Large language models (LLMs) have emerged as strong zero-shot rankers, with the pointwise paradigm remaining the most efficient and parallelizable formulation~\citep{sun2025investigation,liang2022holistic,zhuang2024beyond}. Yet pointwise LLM rankers consistently lag behind on pairwise, setwise and listwise alternatives that explicitly introduce cross-document comparisons~\citep{sun2023chatgpt,pradeep2023rankvicuna}. The standard explanation attributes this gap to the absence of comparative information between documents. We argue that this explanation is incomplete, and a second factor is a calibration gap: \textit{pointwise LLM rankers encode more ranking-relevant signal in their hidden representations than the output head actually reads out at scoring time}.

\begin{figure*}[t]
    \centering
    \includegraphics[width=\linewidth]{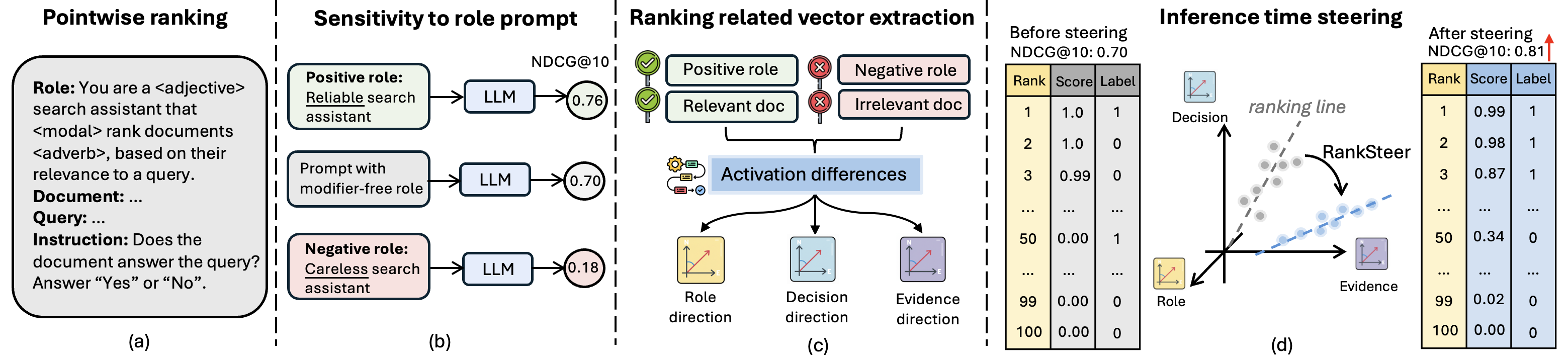}
    \caption{(a)(b) Pointwise LLM ranking prompt and its sensitivity to role phrasings. (c) Decomposing ranking decisions into decision, evidence, and role directions. (d) Performance gain after steering in inference time.}
    \label{fig:figure1}
\end{figure*}
Several existing observations support this view. First, the relevance score in pointwise ranking is produced by a single one-dimensional projection, such as the logit difference between ``Yes'' and ``No''~\cite{liang2022holistic}.
The hidden states feeding this projection, however, are in a space of thousands of dimensions, so information present in the representation but not aligned with the single output projection is discarded at scoring. Second, as shown in Figure~\ref{fig:figure1}(a,b), the brittleness of pointwise rankers to prompt phrasing and role framing~\citep{sun2025investigation,wang2025role} further suggests that the readout, rather than the representation, is the unstable component. Third, mechanistic analyses show that role and content signals propagate through largely separate channels in LLM rankers' hidden states~\citep{wang2025role}, suggesting a structured representation whose components could in principle be addressed independently.

These observations point to a different intervention method than prompt engineering.
Existing attempts to calibrate pointwise scores act only at the input or scoring level~\citep{long2025precise,li2025leveraging,wang2025role} and leave the hidden representation untouched.
Rather than reshaping the input until the output head produces a better score, we investigate whether the representation can be re-aligned so that more of the ranking-relevant signal it already contains is reflected in the score. This shifts the problem from input search to post-hoc, representation-level calibration of a frozen LLM.

We propose RankSteer (Figure~\ref{fig:method}), a modular activation-steering framework for calibrating zero-shot pointwise LLM rankers. RankSteer characterizes ranking behavior through up to three steerable directions in the hidden space: (1) a \textbf{decision direction}, which is the input-agnostic axis along which the output head produces relevance scores;
(2) an \textbf{evidence direction}, which captures the relevance signal present in the hidden states but orthogonal to the decision axis (the calibration gap); and (3) an optional \textbf{role direction}, an orthogonal axis that modulates how decisively the decision direction is applied, which is relevant for base rankers with role-style prompting. At inference time, RankSteer applies simple projection-based interventions along these directions to recalibrate ranking representations, \textbf{without modifying model weights or introducing cross-document comparisons}.

Because the decision direction is induced directly from the model output head, RankSteer is broadly compatible with all pointwise LLM rankers whose relevance score reduces to a two-token contrast, regardless of differences in prompting or scoring design. We instantiate it on two structurally distinct bases, RankSteer-Role and RankSteer-Ref, where both improve over their respective base on the majority of benchmarks using only a handful of out-of-distribution queries, indicating that the calibration gap generalizes across binary-contrast pointwise settings.

Our contributions are: (1) We propose RankSteer, the first post-hoc activation-steering framework for pointwise LLM reranking, which introduces projection-based interventions along decision, evidence, and role directions to steer relevance scoring at the representation level.
(2) We instantiate RankSteer on two distinct pointwise bases, RankSteer-Role and RankSteer-Ref, and show that both improve over their respective base on the majority of ranking benchmarks across three backbones. (3) We provide a geometric analysis showing that calibration operates by concentrating each query's document representations along an existing ranking geometry, and that representation-level calibration recovers more ranking quality than LoRA fine-tuning on the same anchor set. Our code is available at \url{https://github.com/menauwy/ranksteer}.

\section{Related Work}
\paragraph{\textbf{Zero-Shot Ranking.}}\label{related:zeroshot}
Unlike supervised rankers (e.g., MonoT5~\citep{nogueira2020document}, RankT5~\citep{zhuang2023rankt5}) that require substantial labeled data, LLM rankers work in a zero-shot setting. Under the traditional pointwise paradigm, a model scores each (query, document) pair independently~\citep{sachan2022improving,zhuang2023open,liang2022holistic,zhuang2024beyond}.
A line of work aims to improve the reliability of these scores, which we view as \textit{calibrating} the readout. Existing approaches operate at the input or scoring level, including aggregation-level contrastive post-processing such as PAGC~\citep{long2025precise}, reference-level anchoring such as RefRank~\citep{li2025leveraging}, and prompt-level role framing~\cite{wang2025role}. However, these methods do not modify the representation consumed by the output head, leaving the calibration gap insufficiently addressed. RankSteer is the first approach to address calibration at the representation level.
Comparative methods, including pairwise~\citep{qin2024large,luo2024prp}, setwise~\citep{zhuang2024setwise} and listwise~\citep{ma2023zero,pradeep2023rankvicuna,sun2023chatgpt} prompting, improve ranking through explicit cross-document comparison. However, they incur high inference cost, limited parallelism, and positional bias~\citep{long2025precise}. RankSteer instead remains within the pointwise paradigm and avoids cross-document comparison.


\paragraph{\textbf{Role-Play Prompting.}}
Role-play prompting assigns an explicit role or persona to a language model, guiding its behavior and output style~\citep{shanahan2023role}. Modern LLMs exhibit strong role-playing abilities, enabling them to better perform various tasks~\citep{kong2024better,wang2024rolellm}. 
For ranking tasks, LLMs have been shown to be highly sensitive to role-play instructions, heavily affecting their effectiveness~\citep{sun2025investigation,wang2025role}.
Recent causal intervention studies~\citep{meng2022locating,zhang2023towards} further show that role-play instructions are processed largely independently from query-document representations in the model's hidden layers~\citep{wang2025role}. Building on these findings, our work moves beyond prompt-level manipulation by extracting and steering role-related activation directions, enabling post-hoc and interpretable control over ranking behavior.

\paragraph{\textbf{Activation Steering.}}
Activation steering has emerged as a lightweight and transparent paradigm for controlling LLM behavior without modifying model weights~\citep{zou2023representation,bartoszcze2025representation}. The core premise is that many semantic concepts, behaviors, and decision patterns are encoded as approximately linear directions in the activation space, which can be identified and manipulated at inference time~\citep{park2023linear,rimsky2024steering}. Steering directions are typically extracted using contrastive examples, probing classifiers, or sparse auto-encoders, and injected into hidden states to bias model outputs in a targeted manner~\citep{bayat2025steering,wang2025improving,hong-etal-2025-reasoning}.
This paradigm has been successfully applied across a wide range of tasks~\citep{chen2025persona,wang2025adaptiveactivation,li-etal-2025-revisiting,zhao-etal-2025-adasteer,fang2026controllablellmreasoningsparse}. 
Recent work has shown that role-playing behavior can be steered via internal activations using sparse autoencoders, leading to improved \textit{reasoning} performance and highlighting the potential of activation-level control beyond surface-level prompt engineering~\citep{wang2025improving}.
The application of activation steering to ranking remains unexplored. 
We bridge this gap by showing that relevance, decision, and role-play signals in pointwise LLM rankers can be disentangled and calibrated post hoc, enabling effective ranking improvements without modifying model parameters or introducing cross-document comparisons.

\section{Methodology}
\begin{figure*}[ht]
    \centering
    \includegraphics[width=\linewidth]{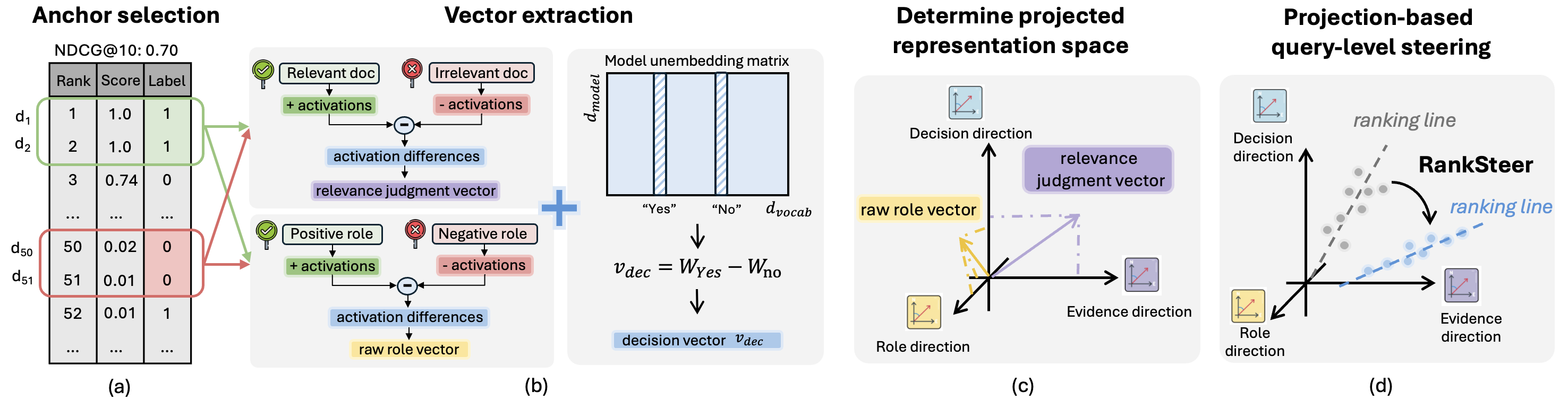}
    \caption{Overview of the RankSteer framework. (a) Selection of query-document anchors for constructing contrastive input pairs. (b) Extraction of ranking-related steering vectors from anchor data. (c) Construction of a projection-based representation space using the extracted vectors. (d) Inference-time, query-level activation steering via projections in the constructed space.}
    \label{fig:method}
\end{figure*}
We propose a post-hoc activation-steering framework that calibrates zero-shot pointwise LLM rankers at the representation level, illustrated in Figure~\ref{fig:method}. For brevity, RankSteer refers to RankSteer-Role throughout this section.
\subsection{Preliminaries}
\paragraph{Pointwise Zero-Shot Reranking.}
\label{sec:pointwise_preliminary}
Following established protocols~\citep{nogueira2020document}, we adopt a binary Yes/No formulation~\citep{liang2022holistic}, which lets us analyze relevance judgments directly through the model's internal representations associated with the final token. Details are in Appendix~\ref{app:pointwiseranking}.
\paragraph{Activation Steering Framework.}
\label{sec:framework}
Activation steering involves two main steps: Steering Vector Extraction and Inference-Time Intervention. Following prior work~\citep{bartoszcze2025representation,zhao-etal-2025-adasteer}, we adopt these preliminaries and refer the reader to Appendix~\ref{app:asframework} for details.

\subsection{Constructing Steering Vectors}
The calibration gap suggests that the hidden state contains more than the output head reads. We decompose the two into a decision direction (what the head reads) and an evidence direction (what it leaves behind), with an optional role direction for role-prompted bases. We construct the steering vectors from a small set of anchor queries from a separate dataset, using them solely for direction estimation without any parameter updates, rather than for fine-tuning.

\subsubsection{Decision Vector}
The decision vector defines a global relevance axis induced by the output head. In pointwise ranking, the model maps a hidden state to the logits of ``Yes'' and ``No'' via its output embedding matrix $W \in \mathbb{R}^{V \times d}$. Letting $W_{\text{yes}}$ and $W_{\text{no}}$ denote the corresponding weight rows, we define the decision vector as their normalized difference:
\begin{equation}
    v_{\text{dec}} = \frac{W_{\text{yes}} - W_{\text{no}}}{|W_{\text{yes}} - W_{\text{no}}|}
\end{equation}
Being input-agnostic, $v_{\text{dec}}$ is shared across all queries and documents and fixes the direction along which the hidden state is projected to form a judgment. Steering along this axis directly affects the model's decisiveness without altering the underlying evidence representations.

\subsubsection{Evidence Vector}
To extract a relevance signal that is consistent with the model's existing ranking behavior with respect to a given dataset, we construct contrastive anchor pairs under both label and ranking constraints. 
Specifically, for a query $q_i$, we select $n$ pairs of documents where the relevant document $d_{i,j}^{+}$ is ranked higher and the irrelevant document $d_{i,j}^{-}$ is ranked lower by the vanilla LLM.
This ensures that the resulting contrast captures how the model currently distinguishes relevance, rather than representations associated with model errors or uncertainty. 
Let $h_{i,j}^{(l)+}$ and $h_{i,j}^{(l)-}$ denote the corresponding last-token hidden states at layer $l$. The raw relevance judgment vector $v_{\text{rel}}^{(l)}$ is computed across $m$ queries as:
\begin{equation}
    v_{\text{rel}}^{(l)} = \frac{1}{m \cdot n} \sum_{i=1}^{m} \sum_{j=1}^{n} \left( h_{i,j}^{(l)+} - h_{i,j}^{(l)-} \right)
\end{equation}
\paragraph{Disentanglement and Calibration.} 
While $v_{\text{rel}}^{(l)}$ encodes relevance, part of this signal is aligned with the decision axis the head reads out, while the majority lies off-axis. 
To isolate the residual relevance evidence that is present in the hidden states but not directly exploited for the final ``Yes/No'' logit, we project $v_{\text{rel}}^{(l)}$ onto the orthogonal complement of the decision axis and normalize it:
\begin{equation}
    v_{\text{evid}}^{(l)} = \text{Norm}(v_{\text{rel}}^{(l)} - \langle v_{\text{rel}}^{(l)}, v_{\text{dec}} \rangle v_{\text{dec}})
\end{equation}
While the decision vector determines whether a document is judged relevant, the evidence vector reflects how strongly document content supports relevance in a way that is comparable across documents. As such, evidence serves as a calibration signal: steering along this direction modulates the contribution of under-utilized relevance cues without altering the underlying decision boundary.

To verify that $v_{\mathrm{evid}}$ is not a noise-dominated residual, we measure $|\cos(v_{\mathrm{rel}}, v_{\mathrm{dec}})|$ across layers, backbones, and bases over the cross-validation anchor folds. The two directions show minimal alignment (mean $|\cos|=0.03$--$0.12$, max $<0.50$), close to the random baseline in $\mathbb{R}^{4096}$ ($\approx 0.013$). Thus, $v_{\mathrm{evid}}$ preserves 98--99\% of $v_{\mathrm{rel}}$'s norm on average, indicating that it retains meaningful signal rather than noise (Appendix~\ref{app:collinearity}).

\subsubsection{Role Vector}
Beyond relevance signals derived from the query-document content, LLMs are highly sensitive to high-level task framing and role-play instructions. Such role signals do not introduce new relevance evidence, but instead influence how the model applies its relevance judgment, acting as a form of behavior-level control. 
We extract a raw role direction from prompt pairs differing only in role instruction (positive vs. negative) with query and document fixed (Appendix~\ref{app:role_play}).

Let $h_{i,j,k}^{(l)\text{role}+}$ and $h_{i,j,k}^{(l)\text{role}-}$ be the hidden states for $t$ contrastive role pairs. The raw role vector $v_{\text{role\_raw}}^{(l)}$ is aggregated over $m$ queries and $n$ documents. 
{\small
\begin{equation}
    v_{\text{role\_raw}}^{(l)} =
    \frac{1}{t} \frac{1}{n} \frac{1}{m}
    \sum_{k=1}^{t} \sum_{i=1}^{n} \sum_{j=1}^{m}
    \left(h_{ijk}^{(l)\text{role}+} - h_{ijk}^{(l)\text{role}-}\right)
\end{equation}
}
We then project it off both relevance axes and normalize:


{\small
\begin{equation}
\begin{split}
    v_{\text{role}}^{(l)} = \text{Norm}\big(
    v_{\text{role\_raw}}^{(l)}
    &- (v_{\text{role\_raw}}^{(l)} \cdot v_{\text{dec}})\, v_{\text{dec}} \\
    &- (v_{\text{role\_raw}}^{(l)} \cdot v_{\text{evid}}^{(l)})\, v_{\text{evid}}^{(l)}
    \big)
\end{split}
\end{equation}
}
After normalization, the resulting $v_{\text{role}}^{(l)}$ is orthogonal to both relevance-related axes. 
We do not attempt to analyze whether role and relevance are naturally disentangled. Instead, we explicitly construct independent directions to isolate the effect of role-play on relevance judgments from relevance evidence itself.

\subsection{Layerwise Steering at Inference Time}
We intervene on the last-token hidden state $h^{(l)}$ at every layer, giving finer control over the model's internal scoring than static prompting.

\paragraph{Projection Onto Steering Directions.}
Given the three unit steering vectors: decision vector $v_{\text{dec}}$, evidence vector $v_{\text{evid}}^{(l)}$, and role vector $v_{\text{role}}^{(l)}$, we first compute scalar projections of the hidden state:
{\small
\begin{equation}
    p_{\text{dec}}^{(l)} = h^{(l)} \cdot v_{\text{dec}}, \quad
    p_{\text{evid}}^{(l)} = h^{(l)} \cdot v_{\text{evid}}^{(l)}, \quad
    p_{\text{role}}^{(l)} = h^{(l)} \cdot v_{\text{role}}^{(l)}.
\end{equation}
}
These projections quantify how strongly the current representation aligns with the three directions at layer $l$.

\paragraph{Decision and Evidence Steering.}
We then apply relevance-oriented steering by adjusting the hidden state along the decision and evidence directions:
\begin{equation}
    \tilde{h}^{(l)} = h^{(l)}
    - \alpha \,\cdot p_{\text{dec}}^{(l)} \, v_{\text{dec}}
    - \beta \,\cdot p_{\text{evid}}^{(l)} \, v_{\text{evid}}^{(l)}
\end{equation}
where $\alpha$ and $\beta$ control the scaling of the decision and evidence axes. 
Steering along $v_{\text{dec}}$ affects the model's global decisiveness, while steering along $v_{\text{evid}}^{(l)}$ modulates under-utilized relevance cues for better cross-document calibration.
Together they shift the representation so that more of its relevance content is reflected in the score, closing the calibration gap by reweighting signals already present rather than injecting external information.

\paragraph{Role-Gated Decision Steering.}
For role-prompted bases, the role signal controls how decisively the decision direction is applied rather than adding relevance evidence. We turn the role projection into a gate and use it to modulate decision steering:
\begin{equation}
    g^{(l)} = \sigma(p_{\text{role}}^{(l)})
\end{equation}
\begin{equation}
    h'^{(l)} = \tilde{h}^{(l)} - \gamma \cdot g^{(l)} \cdot p_{\text{dec}}^{(l)} v_{\text{dec}}
\end{equation}
where $\sigma(\cdot)$ is the sigmoid function and $\gamma$ is the role modulation coefficient. The gate measures the activation strength of role-related features at layer $l$, scaling decision steering accordingly. Role steering adds no relevance information, and only adjusts how decisively the ranker applies its judgment.


\paragraph{Final Scoring.}
After steering all layers, the final hidden state is passed to the output head to compute relevance logits, from which pointwise relevance scores are derived as described in Section~\ref{sec:pointwise_preliminary}. All hyperparameters $\alpha$, $\beta$, and $\gamma$ are fixed at inference time and tuned on an external validation set.


\section{Experiments}
\subsection{Experimental Setup}
\paragraph{Backbone.}
Following prior works on activation steering~\citep{zou2023representation,zhao-etal-2025-adasteer}, we conduct experiments on three representative instruction-tuned decoder-only LLMs as the backbone of our ranker: Llama-3.1-8B-Instruct~\citep{grattafiori2024llama}, Qwen2.5-7B-Instruct~\citep{qwen2025qwen25technicalreport}, and Mistral-7B-Instruct-v0.3~\citep{jiang2023mistral7b, mistral7b_instruct_v0.3}.
\paragraph{Datasets and Metrics.}
We evaluate on two widely used IR benchmarks: TREC Deep Learning (DL)~\citep{craswell2020overview,craswell2021overview} and BEIR~\citep{thakur2021beir}. TREC DL provides graded relevance judgments suitable for constructing reliable anchor data; we therefore use TREC DL 2019 to construct anchor inputs and tune the steering hyperparameters. We evaluate in-distribution generalization on TREC DL 2020, which shares the same corpus as DL 2019 but contains disjoint queries, and out-of-distribution generalization on eight BEIR datasets. We report nDCG@10 as the primary metric, and additionally report MRR@10, MAP and binary accuracy (BA) when analyzing individual steering directions. Binary accuracy measures whether the model correctly predicts relevance by comparing the likelihood of ``Yes'' versus ``No'' with a 0.5 threshold.

\input{table/compare_pointwise}
\paragraph{Baselines.}
Following the calibration taxonomy in Section~\ref{related:zeroshot}, we group pointwise baselines into plain pointwise rankers that read scores directly off the output head (\textbf{QG}~\citep{sachan2022improving}, \textbf{RG-YN}~\citep{liang2022holistic}, \textbf{RG-S(0–k)}~\citep{zhuang2024beyond}) and calibration-oriented rankers that recalibrate at the aggregation level (\textbf{PAGC-YG}~\citep{long2025precise}), reference level (\textbf{RefRank-(k)}~\citep{li2025leveraging}), or prompt level (\textbf{RG-YN\_role}~\citep{wang2025role}). RankSteer calibrates at the representation level and is instantiated on the two strong bases among: \textbf{RankSteer-Role} on RG-YN\_role, applying all three directions and \textbf{RankSteer-Ref} on RefRank-(4), applying only the decision and evidence directions. We further compare against comparative methods (\textbf{PRP}~\citep{luo2024prp}, \textbf{Setwise}~\citep{zhuang2024setwise}, \textbf{RankGPT}~\citep{sun2023chatgpt}) as a different-paradigm reference. Implementation details for all baselines are in Appendix~\ref{app:baselines}.

\paragraph{Implementation Details.}
We retrieve top-100 BM25 candidates with Pyserini~\citep{Lin_etal_SIGIR2021_Pyserini}. Relevance anchors use $m=5$ DL 2019 queries, each with $n=10$ relevant and $n=10$ irrelevant documents aligned with the base ranker's behavior; role-play anchors use $t=3$ instruction pairs from~\citet{wang2025role} (see Appendix~\ref{app:role_play}). The remaining 38 DL 2019 queries tune $\alpha$, $\beta$, $\gamma$, which are then fixed across DL 2020 and BEIR. A multi-fold selection procedure mitigates variance from anchor query sampling. Additional implementation details are provided in Appendix~\ref{app:implementation_details}.

\subsection{Experimental Results}

\subsubsection{Comparison with Pointwise Methods}\label{sec:pointwise_result}
Table~\ref{tab:pointwisecomparison} summarizes nDCG@10 against pointwise baselines. 
We organize our findings around three observations.
\textbf{First, steering improves its base across backbones.} 
Across all three backbones, each steered variant improves over its base on TREC DL 2020, while the gains on BEIR are more mixed rather than uniformly positive, particularly for RankSteer-Role. This suggests that the calibration gap is a general property of binary-contrast pointwise rankers, while the learned steering directions do not transfer equally well across heterogeneous domains. In our strict zero-shot transfer setting, the steering vectors are learned once from TREC DL 2019 and then kept fixed for all BEIR datasets. Consequently, datasets whose document styles or relevance patterns differ more substantially from the source domain may benefit less from the learned directions, and interactions between role prompting and domain characteristics may further contribute to this variation. Nevertheless, RankSteer improves the average performance across the eight BEIR datasets on all three backbones, suggesting that the learned calibration directions retain meaningful transferability without target-domain adaptation.
\textbf{Second, in-distribution gains are stable.} RankSteer-Role achieves the best TREC DL 2020 result on every backbone using only five TREC DL 2019 anchor queries, indicating that effective directions can be extracted from a small anchor set and transferred to unseen in-distribution queries. 
\textbf{Third, the benefits of steering on out-of-distribution datasets are domain-, base-, and backbone-dependent.}
RankSteer-Ref is the more robust variant across BEIR datasets, while RankSteer-Role is less competitive when a plain base is already strong. The datasets on which steering underperforms also vary across backbones, suggesting the extracted directions are partly dataset-dependent: anchors drawn from TREC DL 2019 transfer unevenly to heterogeneous BEIR domains. \textbf{Overall, no single base is uniformly best}, and RankSteer’s modularity enables flexible adaptation, allowing the base method to be chosen according to the backbone model and data distribution.
We provide additional comparisons with non-pointwise baselines in Appendix~\ref{app:non-pointwise_baselines}.

\subsubsection{Ablation of Steering Directions}
\paragraph{Ablation Results.}
We conduct ablation analysis on Llama-3.1-8B-Instruct using a random TREC DL 2019 validation set, as reported in Table~\ref{tab:ablation_study}.
The decision direction improves ranking but lowers binary accuracy, suggesting a rescaling of decision scores, whereas the evidence direction yields smaller ranking gains while leaving binary accuracy unchanged, indicating that it injects under-utilized relevance signal without shifting the decision boundary. Combining the two outperforms either alone, confirming their complementary effects. Adding the role direction gives the best overall ranking while keeping binary accuracy close to the unsteered baseline, suggesting that role steering modulates how decisively the decision direction is applied rather than introducing new relevance evidence. 
\paragraph{Robustness Across Validation Splits.}
While Table~\ref{tab:ablation_study} is based on a single DL19 fold (38 queries), we further evaluated the ablation trends on 8 disjoint DL19 partitions, to verify that the observed trends are not split-specific. Decision steering consistently outperformed the baseline in all 8 folds, and evidence steering improved over decision-only steering in 7/8 folds (paired t-test p=0.0085), confirming the consistency of the observed complementary effects.

\input{table/ablation_study}
\subsubsection{Sensitivity to Anchor Query Set}
\begin{figure}[t]
    \centering
    \includegraphics[width=\linewidth]{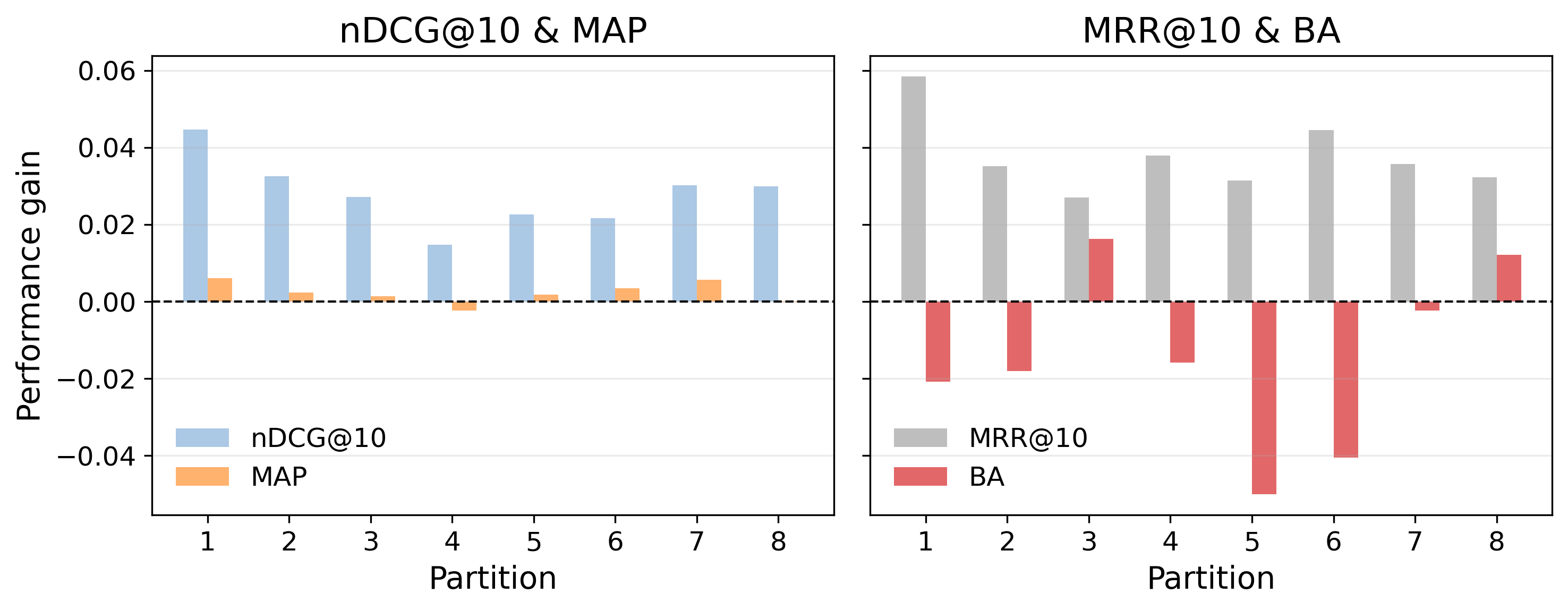}
    \caption{Sensitivity to anchor query set. Performance gain after steering across eight anchor sets (5 queries) of TREC DL 19 using Llama-3.1-8B-Instruct.}
    \label{fig:anchor_sensitivity}
\end{figure}

\begin{figure}[t]
    \centering
    \includegraphics[width=\linewidth]{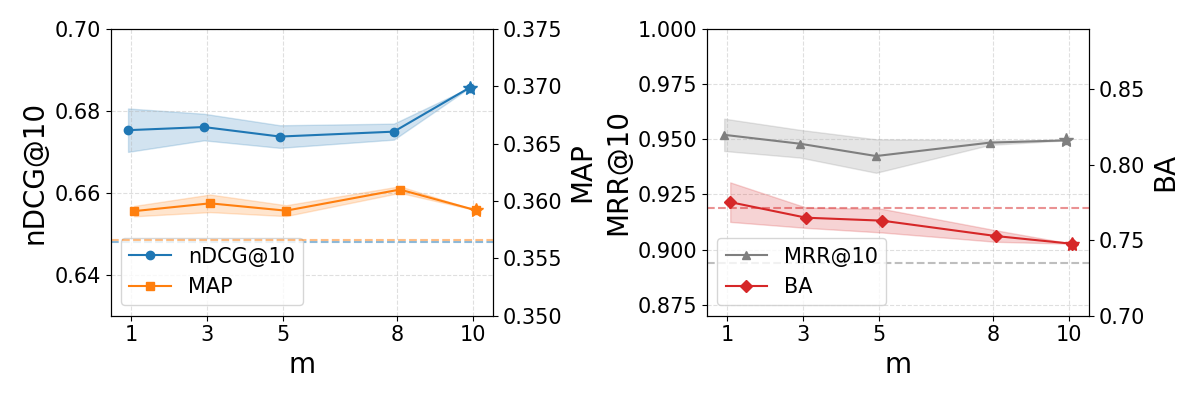}
    \caption{Effect of anchor query counts on TREC DL 2019 with Llama-3.1-8B-Instruct. The horizontal dotted line is the baseline without model steering. For $m=10$, there is only one set of queries and therefore a single measurement.}
    \label{fig:anchor_number}
\end{figure}

We assess robustness to anchor selection along two axes: which queries are used, and how many.
\paragraph{Anchor choice.} We construct steering vectors from eight disjoint five-query partitions of TREC DL 2019, tuning hyperparameters within each (Figure~\ref{fig:anchor_sensitivity}). Steering improves ranking across all eight partitions, with nDCG@10 and MRR@10 increasing by comparable margins in every case. MAP and binary accuracy fluctuate more, but the latter is expected, as steering rescales decision scores and can shift the Yes/No threshold without degrading ranking. The magnitude of improvement varies with the anchor set while its direction stays stable, indicating that RankSteer does not rely on a carefully chosen anchor set.
\paragraph{Anchor count.} We fix 33 queries for validation and sample $m \in \{1, 3, 5, 8, 10\}$ anchors from the remaining 10, repeating five times (Figure~\ref{fig:anchor_number}). Ranking quality is largely insensitive to $m$: even a single anchor is competitive on average, though with noticeably higher variance. We therefore recommend $m=5$ as a balanced setting that delivers robust gains without unnecessary variance.


\begin{figure*}[th]
    \centering
    \includegraphics[width=\linewidth]{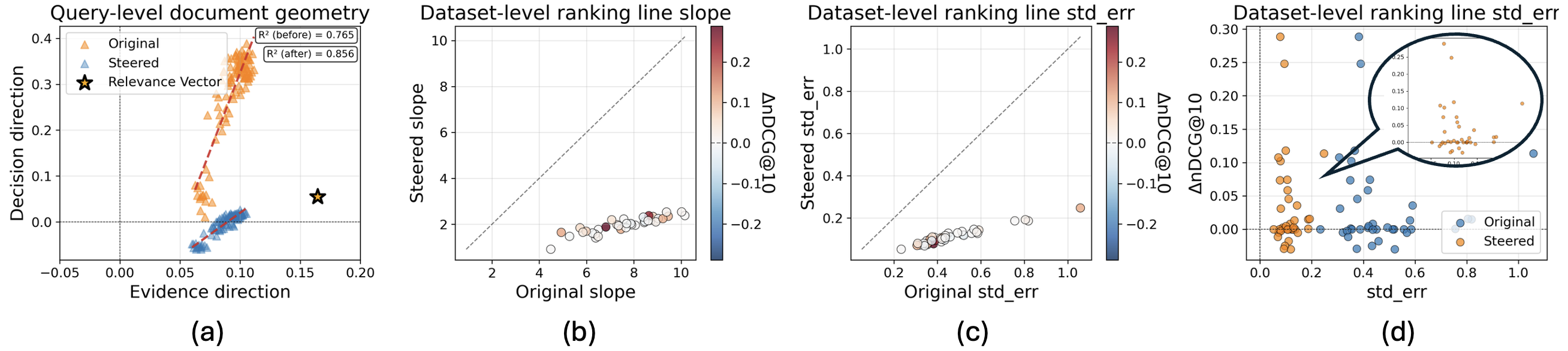}
    \caption{Geometric interpretability of RankSteer-Role at Layer 16 of Llama-3.1-8B-Instruct. (a) Query-level document representations projected onto the decision-evidence plane, illustrating linear ranking geometry. (b) and (c) Dataset-level distributions of the slope and standard error of query-specific ranking lines before and after steering. (d) A complementary view of (c), highlighting how steering redistributes query-level ranking dispersion in relation to changes in nDCG.}
    \label{fig:geometric_plot}
\end{figure*}


\subsubsection{Interpretability from the Geometric Perspective}
Since role-play primarily modulates the decision mechanism rather than document content, we analyze ranking behavior in the two-dimensional space spanned by the decision and evidence directions. Figure~\ref{fig:geometric_plot} visualizes ranking geometry at both the query and dataset levels.

\paragraph{What is being steered?} 
For a given query, document representations lie in an approximately linear arrangement on the decision–evidence plane, which we call a \emph{ranking line} (Figure~\ref{fig:geometric_plot}(a)); this structure persists both before and after steering. Steering therefore, does not create or remove the ranking line but reshapes its geometric statistics: representations become more concentrated, with reduced spread along the decision direction. At the dataset level (Figures~\ref{fig:geometric_plot}(b,c)), for layers where the geometry is well-formed (see full layer geometry in Appendix~\ref{app:full_geo}), steering consistently reduces both the slope and the spread of fitted ranking lines. We quantify this spread using \emph{dispersion}, defined as the standard error of the fitted ranking line. This means steering lowers the dominance of the decision direction relative to evidence, letting representations align more closely with evidence-supported relevance and making ranking structure more coherent across queries. Consistently, the dispersion $\Delta$nDCG relationship in Figure~\ref{fig:geometric_plot}(d) shows that steering shifts queries toward lower dispersion within an unchanged overall pattern, suggesting that performance gains arise from tightening an existing ranking structure rather than introducing a fundamentally different ranking behavior.

\paragraph{Is the geometric change an artifact?}
\input{table/geometry_control}
To verify that the observed geometric changes are not merely a consequence of the projection-and-subtraction operation itself, we perform two control experiments under the same RankSteer setting. First, \emph{Random Direction} applies the identical intervention procedure and coefficients, but replaces the learned steering directions with random unit vectors. Second, \emph{Variance-matched Compression} directly compresses representations toward their mean without using any learned direction, while matching (or exceeding) the overall representation variance reduction induced by RankSteer. Results are summarized in Table~\ref{tab:geometry_control}.
Both controls fail to reproduce the ranking improvement despite applying comparable geometric transformations. Random Direction leaves both ranking-line dispersion and performance nearly unchanged, indicating that the intervention operation alone is insufficient. Meanwhile, Variance-matched Compression reduces global representation variance even more than RankSteer, but produces little change in ranking-line dispersion and no ranking improvement. These results suggest that the observed gains are not caused by generic representation compression, but depend on the learned ranking geometry encoded by the extracted steering directions.

\paragraph{Comparison with fine-tuning}
\input{table/fine_tuned}
\begin{figure}[t]
    \centering
    \includegraphics[width=\linewidth]{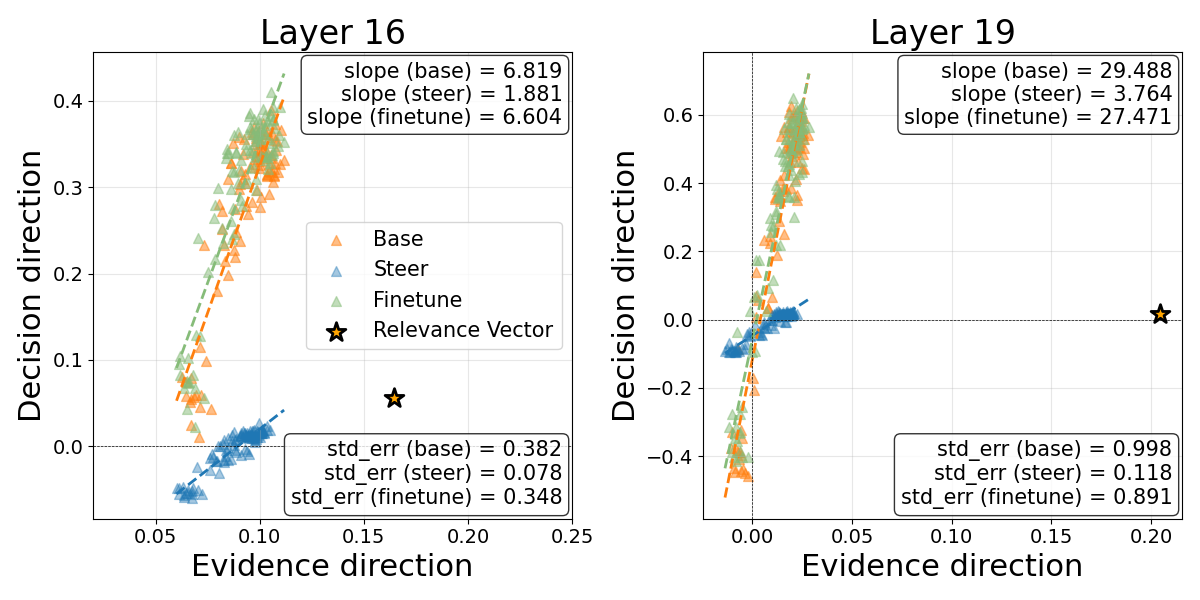}
    \caption{Comparison of query-level document representations under steering and fine-tuning.}
    \label{fig:finetune_plot}
\end{figure}
Using the same five anchor queries, we fine-tune Llama-3.1-8B-Instruct with LoRA~\citep{hu2022lora} and evaluate on the same validation set (Table~\ref{tab:fine_tuned}). With identical supervision data, fine-tuning improves nDCG@10 only modestly, whereas steering with these anchor queries (without finetuning) achieves a markedly larger gain while matching fine-tuning on MRR@10 and without updating any weights; binary accuracy decreases slightly, consistent with the rescaling of decision scores. Geometrically (Figure~\ref{fig:finetune_plot}), fine-tuning largely preserves the base model's ranking lines with only moderate reductions in slope and dispersion, while steering induces a stronger effect, compressing both more substantially and yielding tighter clustering along the decision direction while staying close to the original ranking manifold. 
These results highlight a sample-efficiency advantage in low-resource settings: with only five anchor queries, directly manipulating activation geometry recovers under-utilized ranking capacity more effectively than parameter updates when supervision is scarce.

\section{Conclusion}

We introduced RankSteer, a post-hoc activation-steering framework that calibrates pointwise LLM rankers at the representation level. Rather than reshaping the input, RankSteer recalibrates the representation at inference time through projection-based interventions along a decision direction, an evidence direction, and an optional role direction, without updating model weights or introducing cross-document comparisons. Across three backbones, we apply RankSteer on two strong base methods and see that it improves over its base on most TREC DL and the BEIR datasets, using steering vectors extracted from only five anchor queries. A geometric analysis shows that it works by pulling each query's document representations into a tighter, more consistent spread along an existing ranking geometry. Our results suggest that the calibration gap is a general property of binary-contrast pointwise rankers, and that representation-level calibration offers a favorable trade-off, recovering under-utilized ranking capacity while preserving the full parallelism and efficiency of standard pointwise ranking.

\section*{Acknowledgments}
This publication is part of the project LESSEN with project number NWA.1389.20.183 of the research program NWA ORC 2020/21 which is (partly) financed by the Dutch Research Council (NWO).

\section*{Limitations}
First, our method applies to pointwise rankers and does not directly extend to other paradigms. This boundary follows from how the decision direction is constructed: it is read off the output head as a two-token margin, so RankSteer applies to any pointwise base whose score reduces to such a contrast, covering both the Yes/No base RG-YN\_role and the A/B base RefRank, but not to listwise or setwise methods that lack a single two-token scoring margin. As the first work to bring activation steering to ranking, we leave these paradigms to future work.
Second, we limited our experiments to models of 7B and 8B parameters, leaving other model sizes to future work.
Third, for the role direction we adopt the role-play templates of prior work~\citep{wang2025role} rather than optimizing them, as template design is not a focus of this paper. That the role direction is already effective under these off-the-shelf templates suggests our gains do not depend on carefully engineered role prompts, though a dedicated search over role templates may yield further improvements.

\section*{Ethics Statement}
All models and datasets used in this work are publicly available for research purposes and are used for their intended purpose of evaluating zero-shot ranking. Our use is consistent with their respective licenses and access conditions. We release our method and code for research use only, consistent with the access conditions of the underlying artifacts.

\bibliography{reference}

\appendix

\section{Preliminaries}

\subsection{Pointwise ranking method}\label{app:pointwiseranking}
Given a query $q$ and a set of candidate documents $\mathcal{D}=\{d_1, d_2, \dots, d_n\}$, the model is prompted with a binary instruction to judge the relevance of each pair $(q, d)$, such as: ``Does the passage answer the query? Answer `Yes' or `No'.'' Let $Z_{\text{yes}}$ and $Z_{\text{no}}$ denote the raw logits assigned to the tokens ``Yes'' and ``No'' respectively, at the final generation position. The relevance score $s(q, d)$ is defined as the softmax-normalized probability of the token ``Yes'':
\begin{equation}
    s(q,d) = \frac{\exp(Z_{\text{yes}})}{\exp(Z_{\text{yes}}) + \exp(Z_{\text{no}})}
\end{equation}
The candidate documents are then ranked in descending order of $s(q,d)$.

\subsection{Activation Steering Framework} \label{app:asframework}
\textbf{Steering Vector Extraction.} The steering process begins by identifying a steering vector $v^{(l)} \in \mathbb{R}^d$ that captures a desired attribute at layer $l$. A common practice involves using contrastive prompt pairs. For a given concept (e.g., relevance judgment), let $X^+$ be a set of positive examples (query and relevant documents) and $X^-$ be a set of negative examples (query and irrelevant documents). The steering vector is computed as the mean difference between the hidden states $h_i^{(l)}$ of these pairs:
\begin{equation}
v^{(l)} = \frac{1}{n} \sum_{i=1}^{n} \left( h_{i+}^{(l)}- h_{i-}^{(l)} \right)
\end{equation}
To ensure numerical stability and comparability across different layers, each steering vector is normalized to unit length such that $\|v^{(l)}\|_2 = 1$. In the context of pointwise LLM ranking, we start with a set of vectors $\mathcal{V} = \{v_{relevance}, v_{role}\}$. Here, $v_{\text{rel}}$ represents relevance-judgment-related representation differences, while $v_{\text{role}}$ captures role-induced behavioral biases.\\
\textbf{Inference-Time Intervention.}
In this work, we focus on steering the last token hidden state, which typically aggregates the semantic information of the input sequence. For a transformer model with $L$ layers, the intervention is applied at every layer during the forward pass. 
A basic form of activation steering modifies the hidden state by adding a steering vector scaled by a coefficient $\alpha$:
\begin{equation}
    \tilde{h}^{(l)} = h^{(l)} + \alpha \cdot v^{(l)}
\end{equation}
While the steering vector is identified independently for each layer, its effect is accumulated as the modified signal propagates through the residual stream. A positive $\alpha$ shifts the model's internal state toward the target concept, thereby modulating the output logits toward the desired distribution. More generally, activation steering can be implemented through projection-based interventions, allowing fine-grained control over how hidden states are modulated along specific vectors.

\section{Collinearity Diagnostic between $v_{\mathrm{rel}}$ and $v_{\mathrm{dec}}$}
\label{app:collinearity}

To examine whether $v_{\mathrm{evid}}$ preserves relevance-related information rather than being dominated by projection artifacts, we measure the collinearity between $v_{\mathrm{rel}}$ and $v_{\mathrm{dec}}$. For each backbone (Llama-3.1-8B, Qwen2.5-7B, and Mistral-7B) and base variant (RankSteer-Ref and RankSteer-Role), we compute $|\cos(v_{\mathrm{rel}}, v_{\mathrm{dec}})|$ at each layer, averaged over the cross-validation anchor folds. We additionally measure the fraction of $v_{\mathrm{rel}}$'s norm preserved after removing the $v_{\mathrm{dec}}$ component:
\[
\frac{\|v_{\mathrm{rel}}-\mathrm{proj}(v_{\mathrm{rel}},v_{\mathrm{dec}})\|}{\|v_{\mathrm{rel}}\|}
= \sqrt{1-\cos^2}.
\]

Table~\ref{tab:collinearity} reports the results averaged over all layers. Across all six settings, $v_{\mathrm{rel}}$ and $v_{\mathrm{dec}}$ are far from collinear:
mean $|\cos|$ ranges from 0.03 to 0.12, well above the expected value for two random directions in $\mathbb{R}^{4096}$ ($\approx 0.013$). Consequently, the residual $v_{\mathrm{evid}}$ retains 98--99\% of $v_{\mathrm{rel}}$'s norm on average, indicating that projecting out $v_{\mathrm{dec}}$ removes only a small, decision-aligned component rather than discarding most of the relevance signal.

\begin{figure}[t]
\centering
\includegraphics[width=\linewidth]{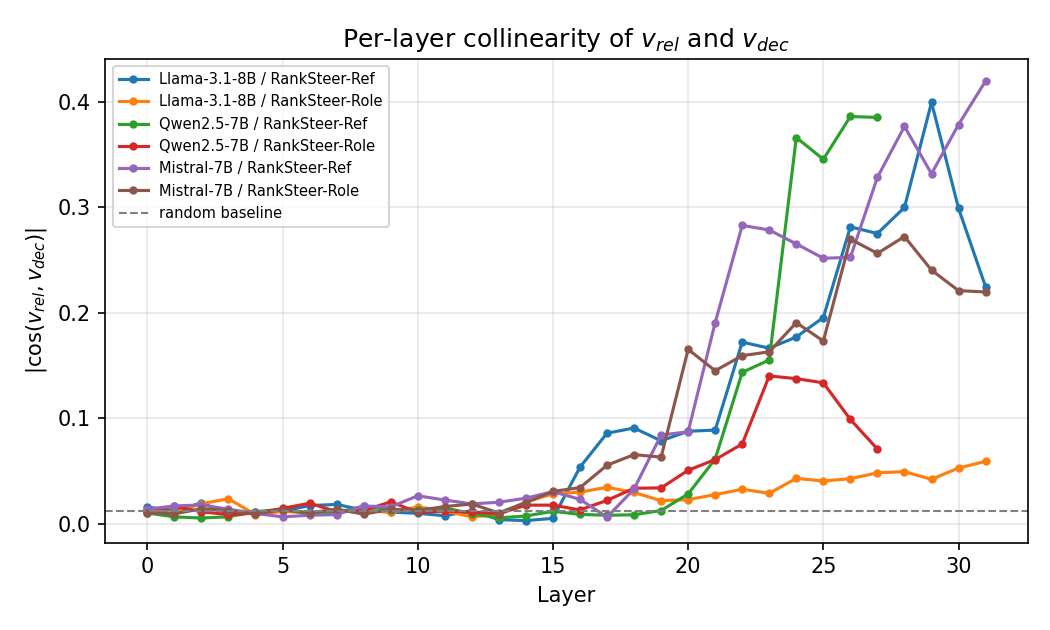}
\caption{Per-layer $|\cos(v_{\mathrm{rel}}, v_{\mathrm{dec}})|$ for all
backbone/base combinations, averaged over cross-validation anchor folds. }
\label{fig:collinearity}
\end{figure}

Figure~\ref{fig:collinearity} presents the layer-wise trend. The alignment remains close to the random baseline in early layers and increases gradually toward later layers, but never approaches complete collinearity in any backbone or layer. This pattern suggests that although the decision direction becomes more identifiable in deeper layers, most of the relevance representation remains distributed outside the single axis read by the output head, supporting the calibration-gap interpretation.

\input{table/collinearity}

\section{Experimental details}
\label{sec:appendix}
\subsection{Role-play prompt}
\label{app:role_play}
Following \citet{wang2025role}\footnote{Available at \url{https://github.com/menauwy/roleplay_zeroshot}}, 
we construct contrastive role-play prompts that share an identical task structure and differ only in three role-defining tokens (highlighted in \textcolor{red!70!black}{red}): an \emph{adjective} describing the model's capability to perform the task, a \emph{modal verb}, and an \emph{adverb} guiding how the model executes the task. We design two role variants, \emph{positive} and \emph{negative}, which are intended to elicit good and degraded ranking behavior, respectively, providing a controlled behavioral contrast for steering vector extraction.
\begin{tcolorbox}[title=Role-Play Prompt Template for Steering Vector Construction, colback=gray!5, colframe=black!50]
You are a \textcolor{red!70!black}{\{adjective\}} search assistant that \textcolor{red!70!black}{\{modal verb\}} rank passages \textcolor{red!70!black}{\{adverb\}}, based on their relevance to a query. \\
Passage: \{passage\} \\
Query: \{query\} \\
Does the passage answer the query? \\
Answer `Yes' or `No'. \\
Answer: 
\end{tcolorbox}

To ensure that each backbone exhibits a clear behavioral gap between the two role types, we adopt the model-specific lexical selections released by \citet{wang2025role}, which were empirically verified 
to yield the expected behavioral contrast for each model. 
Table~\ref{tab:role_tokens} lists the three positive and three negative role prompts used for each of our backbones (Llama-3.1-8B-Instruct, Qwen2.5-7B-Instruct, and Mistral-7B-Instruct-v0.3).
\input{table/role_tokens}

\subsection{Implementation Details.}
\label{app:implementation_details}
For each of the $m=5$ relevance anchor queries from TREC DL 2019, we take $n=10$ relevant documents from the top of the base ranker's output and $n=10$ irrelevant documents starting from rank 50. Choosing relevant documents from the top keeps the extracted vectors aligned with the model's own ranking behavior, while starting irrelevant selection at rank 50 avoids overly diverse documents that would weaken the extracted direction.
Because anchor queries are randomly sampled from TREC DL 2019, we mitigate the resulting variance with a multi-fold selection procedure. We split DL 2019 into 9 partitions, where each of the first 8 partitions provides a distinct set of 5 anchor queries. For each partition, we compute the corresponding anchor vector and evaluate it on the validation queries. The vector achieving the highest validation gain is used in all subsequent experiments.
The tuned steering coefficients $(\alpha,\beta,\gamma)$ are $(0.60,0.16,0.04)$ for Llama, $(0.25,-0.06,0.08)$ for Qwen, and $(0.40,0.00,0.06)$ for Mistral.
All experiments are conducted on a workstation equipped with four NVIDIA L40S GPUs (48GB memory each).

\section{Baseline and Method Variant Details}
\label{app:baselines}

\subsection{Pointwise Baselines}\label{app:pointwise_baselines}
\textbf{QG}~\citep{sachan2022improving} scores documents based on query generation likelihood.
\textbf{RG-YN}~\citep{liang2022holistic} uses the probability of generating ``Yes'' relative to ``No'' as the relevance score.
\textbf{RG-S(0–k)}~\citep{zhuang2024beyond} extends RG-YN by prompting the model to rate relevance on a discrete scale; following prior work, we set $k=4$.
\textbf{PAGC-YG}~\citep{long2025precise} is a post-aggregation technique combining GCCP (global-consistent pairwise comparisons in a pointwise framework) with RG-YN.
\textbf{RefRank-(4)}~\citep{li2025leveraging} reformulates pointwise ranking as an A/B comparison: given a query and a candidate passage, the model compares the candidate against a fixed set of top $K$ ranked reference passages and aggregates the results. We set $k=4$ following the best practice of the paper.
\textbf{RG-YN\_role}~\citep{wang2025role} augments RG-YN with a neutral role-play description from appendix~\ref{app:role_play} but with the three role-defining tokens (adjective, modal, adverb) removed.

\input{table/compare_other}
\subsection{Comparative Baselines}
\textbf{PRP}~\citep{luo2024prp} produces relative relevance labels for document pairs and aggregates them into a ranking via heapsort, returning the top $k=10$. \textbf{Setwise}~\citep{zhuang2024setwise} replaces each pairwise judgment with a setwise comparison over a small group of passages, using generation-based scoring and heapsort with branching factor \texttt{num\_child}=2, a single permutation, and top $k=10$. \textbf{RankGPT}~\citep{sun2023chatgpt} generates a global ordering via sliding-window listwise prompting (window size 4, step size 2) over the full candidate list, repeated for 5 passes with generation-based scoring.

\subsection{RankSteer-Ref: Implementation Details}
\label{app:ranksteer_ref}
RankSteer-Ref is built on the reference-based RefRank~\citep{li2025leveraging} base and follows the same activation-steering recipe as RankSteer but drops the role-play dimension and keeps only two: a decision direction and an orthogonalised evidence direction. Because RefRank scores candidates with a pairwise prompt:
\begin{tcolorbox}[title=Prompt Template for RefRank and RankSteer-Ref, colback=gray!5, colframe=black!50]
Given a query \{query\}, which of the following two passages is more relevant to the query? \\
A: \{passage\} \\
B: \{reference passage\} \\
Output A or B:
\end{tcolorbox}
the decision direction is taken as the A-vs-B unembedding margin from the LM head rather than the pointwise yes/no margin, and anchors are built in the same A/B layout so the geometry aligns with inference. The evidence vector is extracted once using the BM25 top-1 reference and is then reused across all references when RefRank runs in multi-reference (top-$k$) mode.

\section{Effectiveness and efficiency with non-pointwise methods}\label{app:non-pointwise_baselines}
\subsection{Effectiveness comparison}
Table~\ref{tab:othercomparisons} compares RankSteer with representative pairwise, setwise, and listwise ranking methods, and BM25.

While comparative methods such as pairwise and listwise prompting generally achieve stronger effectiveness by explicitly introducing document comparisons, RankSteer operates under a strictly pointwise setting without access to cross-document information. 
The relevant question is how much of the gap can be closed by post-hoc calibration alone.
As a general observation, RankSteer is significantly but only marginally behind listwise and the gap to setwise is not statistically significant.
In particular, RankSteer achieves performance comparable to strong comparative methods (listwise) on TREC DL 2020 for all backbones. 
On BEIR datasets, where the data is out-of-distribution, RankSteer on average stays close to setwise on both Llama (RankSteer-Ref 0.4600 vs. 0.4608) and Qwen (RankSteer-Ref 0.4814 vs. 0.4830).
On some BEIR datasets, RankSteer even attains the best score of any method, such as robust04 and nfcorpus on Llama. 
The two instantiations behave differently, consistent with Section~\ref{sec:pointwise_result}. RankSteer-Ref is the more robust variant against comparative baselines. RankSteer-Role is competitive only when its role-prompted base is strong. 
\subsection{Efficiency comparison}
From a computational perspective, RankSteer retains the same lowest time complexity as standard pointwise ranking, introducing only a fixed overhead for activation steering.
Specifically, RankSteer introduces no additional forward passes, no extra generated tokens, and no additional context length. It only applies lightweight projection operations on the hidden states during the existing forward pass. In our measurements on Llama-3.1-8B, this translates to only \textbf{+2.6\%} latency for the decision+evidence variant (RankSteer-Ref) and \textbf{+5.5\%} latency for the full three-direction variant (RankSteer-Role), with no measurable increase in peak GPU memory. The additional computation is therefore a fixed per-(query, document) cost and remains independent of corpus size, ranking depth, or the number of queries.

In contrast, comparative methods require additional document interactions, leading to $O(N)$ complexity for listwise approaches and $O(\log N)$ for heap-based pairwise and setwise methods.\footnote{\cite{long2025precise} distinguish between time complexity and number of LLM calls in the analysis of the complexity of ranking methods. Here we focus on time complexity.}
Together, these results suggest that a significant portion of the performance gap between pointwise and comparative methods stems from how internal relevance signals are utilized and scaled, rather than from the absence of explicit cross-document comparisons. RankSteer therefore offers a favorable trade-off, recovering much of this gap while preserving the simplicity, efficiency, and full parallelism of pointwise inference.

\section{Query-level document geometry}\label{app:full_geo}
\begin{figure*}[th]
    \centering
    \includegraphics[width=0.8\linewidth]{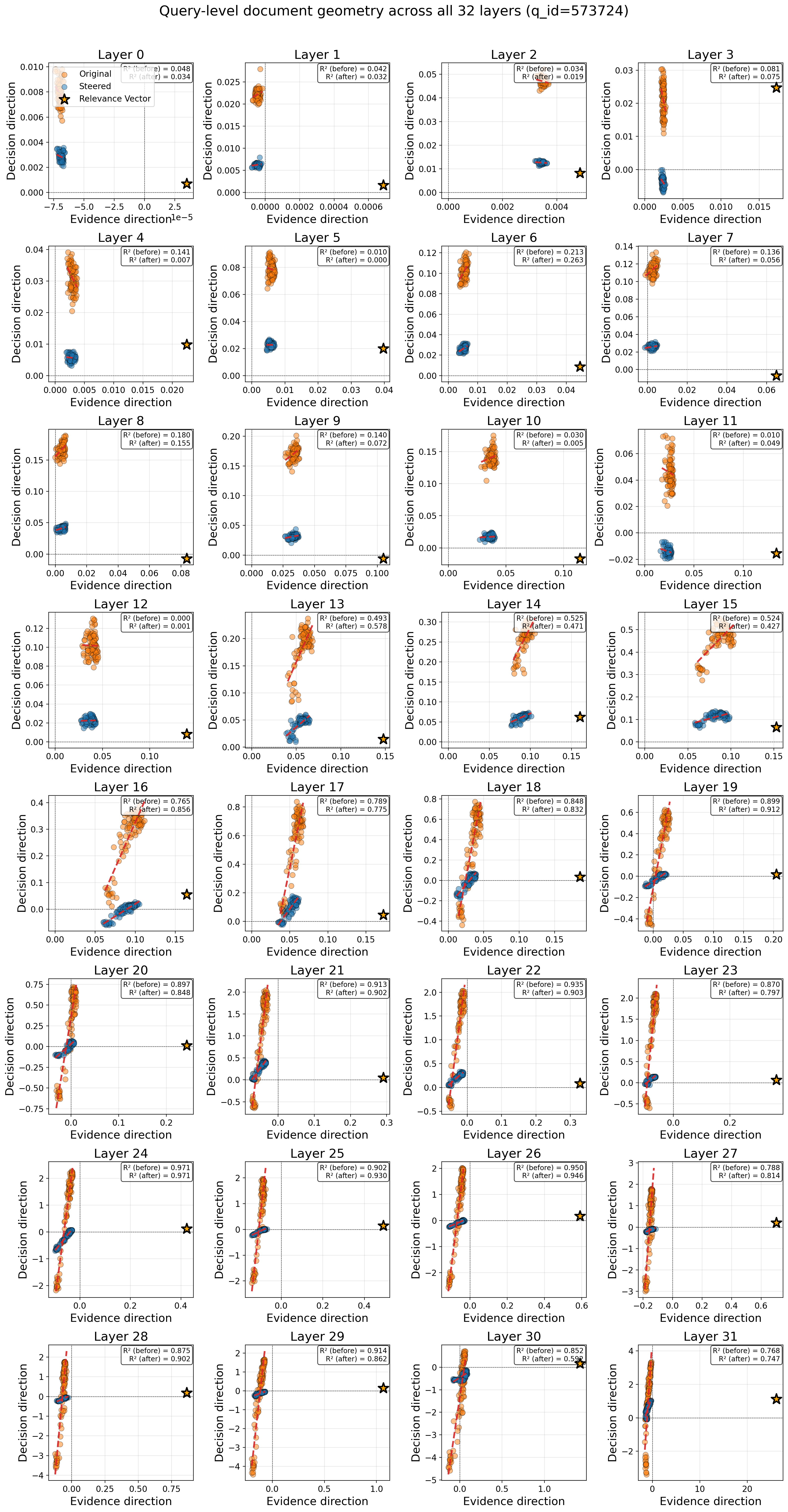}
    \caption{Query-level document geometry across all 32 layers (q = 573724, TREC DL-2019) using RankSteer-Role.}
    \label{fig:full_geometry}
\end{figure*}
In Figure~\ref{fig:full_geometry}, we present the layer-wise steering effect on the document representations of a random query from TREC DL 2019, using RankSteer-Role on Llama-3.1-8B-Instruct. Across layers, we observe three consistent patterns.
First, the \textit{ranking line}, a single linear axis along which all 100 candidate documents organize in the (Evidence, Decision) plane, is absent in early layers. There, the projections form an isotropic cloud with low slope and low $R^2$. The line emerges around L13 and is stably preserved from roughly L16 onward, indicating that ranking is a mid-to-late-layer geometric phenomenon implemented in the residual stream.
Second, after steering, the slope of this line consistently decreases at nearly every well-defined layer. This is the signature of RankSteer: it reduces the ranking's reliance on the raw decision readout, so that the ordering depends less steeply on it while the documents remain on the line.
Third, despite the decrease of slope, $R^2$ is largely retained, and is slightly improved in several mid layers. This means steering re-weights documents within the Evidence–Decision subspace without introducing variance orthogonal to the ranking line.

\end{document}

%% file: table/compare_pointwise.tex
\begin{table*}[ht]
  \centering
  \scriptsize
  \setlength{\tabcolsep}{4pt}
  \begin{tabular}{c|c|c|cccccccc|c}
    \toprule
     & & \textbf{TREC DL} & \multicolumn{9}{c}{\textbf{BEIR}} \\
    \midrule
     & \textbf{Methods} & \textbf{dl20} & \textbf{covid} & \textbf{touche} & \textbf{signal} & \textbf{news} & \textbf{scifact} & \textbf{robust04} & \textbf{dbpedia} & \textbf{nfcorpus} & \textbf{Avg} \\
    \midrule
     & BM25 & 0.4796 & 0.5947 & 0.4422 & 0.3304 & 0.3952 & 0.6789 &0.4070 & 0.3180 & 0.3218 & 0.4360\\
    \midrule
    & QG & 0.3498 & 0.6925 & 0.2205 & 0.1015  & 0.4221 & 0.6041  & 0.4508 & 0.2817& 0.3506& 0.3905\\
    & RG-YN & 0.6109 & 0.7261 & 0.2264 & 0.2482 & 0.3665 & 0.5677 & 0.5046 & 0.3093 & 0.3563 &0.4131\\
    & RG-S(0,4) & 0.5190 & \underline{0.7902} & 0.2205 & 0.2637 &0.4451 &0.6241 & 0.5039 & \underline{0.3666} & 0.3646 & 0.4473\\
    \cmidrule(lr){2-2}\cmidrule(lr){3-3}\cmidrule(lr){4-12}
    & PAGC-YG & 0.6193 & 0.7534 & 0.2424 & 0.2586 & \underline{0.4503} & 0.6485 & 0.5199 &0.3420 & 0.3596 & 0.4468\\
    \rowcolor{gray!7} \cellcolor{white}& RefRank-(4) & 0.5576 & 0.7828 & 0.2418 & 0.2720 & 0.4408 & 0.6443 & 0.5075 & 0.3483 & 0.3487 & 0.4203 \\
    \rowcolor{blue!4} \cellcolor{white}& RG-YN\_role & \underline{0.6230} & 0.7792 & 0.2227 & 0.2661 & 0.4456 & \textbf{0.6606} & \textbf{0.5321} & 0.3574 & \underline{0.3715}&0.4544 \\
    \cmidrule(lr){2-2}\cmidrule(lr){3-3}\cmidrule(lr){4-12}
    \rowcolor{gray!14} \cellcolor{white} & RankSteer-Ref & 0.5936$^\dagger$ & \textbf{0.7942}$^\dagger$ & \underline{0.2443} & \textbf{0.2893}$^\dagger$ & \textbf{0.4660}$^\dagger$ & \underline{0.6561}$^\dagger$ & 0.5126 & 0.3598$^\dagger$&0.3575$^\dagger$ & \textbf{0.4600} \\
    \rowcolor{blue!8} \cellcolor{white} \multirow{-8}{*}{\rotatebox{90}{Llama-3.1-8B}}  & RankSteer-Role & \textbf{0.6461}$^\dagger$ & 0.7863 & \textbf{0.2445}$^\dagger$ & \underline{0.2779} & 0.4462 & 0.6498 & \underline{0.5207} & \textbf{0.3752}$^\dagger$ & \textbf{0.3732} & \underline{0.4588}\\
    \midrule
    & QG & 0.4287 & 0.6621 &0.2277 &0.1010 & 0.3048 &0.4816 &0.3506 &0.3075 & 0.3019& 0.3422\\
    & RG-YN & 0.4897 & 0.7315 & 0.1976 & 0.2243  &0.3963 &0.5143 & 0.4760 &0.2791 & 0.3335 & 0.3941 \\
    & RG-S(0,4) &0.5491 & \underline{0.7892} & 0.2468 & 0.2679 & \textbf{0.4678}  & \textbf{0.6324} & 0.4856 & \underline{0.3672} &\underline{0.3512} & \textbf{0.4510}\\
    \cmidrule(lr){2-2}\cmidrule(lr){3-3}\cmidrule(lr){4-12}
    & PAGC-YG & \underline{0.5522} & 0.7775 & 0.2212 & 0.2586  &0.3949 & 0.4476 & \underline{0.4894} &0.3307 & 0.3338 & 0.4067 \\  
    \rowcolor{gray!7} \cellcolor{white}& RefRank-(4) & 0.5290 & 0.7520 & \underline{0.2885} & \underline{0.2858} & 0.4382 & 0.5987 & 0.4823 & 0.3454 & 0.3383 & 0.4186\\
    \rowcolor{blue!4} \cellcolor{white}& RG-YN\_role & 0.5082 & 0.7136 & 0.1998 & 0.2268 & 0.4192 & 0.5304 & 0.4884 & 0.2954 & 0.3401 & 0.4017\\
    \cmidrule(lr){2-2}\cmidrule(lr){3-3}\cmidrule(lr){4-12}
    \rowcolor{gray!14} \cellcolor{white} & RankSteer-Ref & 0.5407$^\dagger$ & 0.7729$^\dagger$ & \textbf{0.2965} & 0.2849 & 0.4432 & \underline{0.6127}$^\dagger$ & 0.4892 & 0.3375 & 0.3422 & 0.4474\\
    \rowcolor{blue!8} \cellcolor{white}  \multirow{-8}{*}{\rotatebox{90}{Mistral-7B}} & RankSteer-Role & \textbf{0.6355}$^\dagger$ & \textbf{0.8019}$^\dagger$ & 0.2254$^\dagger$ & \textbf{0.2877}$^\dagger$ & \underline{0.4441}$^\dagger$ & 0.5664$^\dagger$ & \textbf{0.5123}$^\dagger$ & \textbf{0.3965}$^\dagger$ & \textbf{0.3623}$^\dagger$ & \underline{0.4496} \\
    \midrule
     
    & QG & 0.3982 & 0.6463 & 0.2190 & 0.0922 & 0.2533 & 0.3499 & 0.3617 & 0.2530  & 0.2845& 0.3075\\
    & RG-YN & 0.6203 & 0.6780 & 0.2025 & 0.2305 & 0.3040 & 0.2395 & 0.3484 &0.2733 & 0.3106 & 0.3234\\
    & RG-S(0,4) & 0.6059 & 0.7814 & 0.2333 & 0.2661  & 0.4490 & 0.6261 & 0.4756 & 0.3618 & 0.3599 &0.4441 \\
    \cmidrule(lr){2-2}\cmidrule(lr){3-3}\cmidrule(lr){4-12}
    & PAGC-YG & \underline{0.6500} & 0.7395 & 0.2413 & 0.2761 & 0.3772 & 0.4632 & 0.4488 & 0.3525 & 0.3552 & 0.4067\\
    \rowcolor{gray!7} \cellcolor{white}& RefRank-(4) & 0.6327 & \underline{0.7818} & \textbf{0.2816} & \textbf{0.3046} & \underline{0.4793} & \underline{0.6737} & \underline{0.5142} & \underline{0.4070} & \underline{0.3736} & \underline{0.4489} \\
    \rowcolor{blue!4} \cellcolor{white} & RG-YN\_role & 0.6156 & 0.7212 & 0.1985 & 0.2249 & 0.3048 & 0.2398 & 0.3691 & 0.2909 & 0.3183 & 0.3334 \\
    \cmidrule(lr){2-2}\cmidrule(lr){3-3}\cmidrule(lr){4-12}
    \rowcolor{gray!14} \cellcolor{white} & RankSteer-Ref & 0.6475$^\dagger$ & \textbf{0.7925}$^\dagger$ & \underline{0.2746} & \underline{0.3017} & \textbf{0.4913}$^\dagger$ & \textbf{0.6778} & \textbf{0.5226}$^\dagger$ & \textbf{0.4148}$^\dagger$ & \textbf{0.3759} & \textbf{0.4814} \\
    \rowcolor{blue!8} \cellcolor{white} \multirow{-8}{*}{\rotatebox{90}{Qwen2.5-7B}} & RankSteer-Role & \textbf{0.6594}$^\dagger$ & 0.7678$^\dagger$ & 0.2356$^\dagger$ & 0.2793$^\dagger$ & 0.3807$^\dagger$ & 0.4133$^\dagger$ & 0.4338$^\dagger$ & 0.3749$^\dagger$ & 0.3697$^\dagger$ & 0.4069\\
    \bottomrule
  \end{tabular}
  \caption{Comparison of RankSteer to pointwise ranking baselines in terms of nDCG@10. Within each backbone, methods are ordered as plain pointwise rankers, calibration-oriented rankers, and RankSteer. RankSteer-Role (our main method) applies three-direction activation steering on the RG-YN\_role base. RankSteer-Ref (modular extension) applies decision and evidence steering only on RefRank-(4) as base. Best and second-best results are marked with \textbf{boldface} and \underline{underline} respectively. All LLMs are instruction-tuned versions. $\dagger$ indicates a statistically significant improvement over the
corresponding base under a two-tailed paired $t$-test with $p<0.05$.
  }
  \label{tab:pointwisecomparison}
\end{table*}

%% file: table/ablation_study.tex
\begin{table}[t]
  \centering
\scriptsize
  \setlength{\tabcolsep}{4pt}
  \begin{tabular}{c|ccc|c}
    \toprule
    Vector & nDCG & MRR & MAP & BA\\
    \midrule
     - & 0.6800 & 0.8985 & 0.3656 & 0.7665 \\
    \midrule
    $v_{dec}$ & 0.6963 & 0.9193 &0.3673 & 0.7457\\
    $v_{evid}$&0.6843 & 0.9020 &0.3662 &0.7665\\
    \midrule
    $v_{dec} + v_{evid}$ & \underline{0.7061} & \underline{0.9386}& \underline{0.3701}&0.7625\\
    $v_{dec} + v_{role}$ &0.7029&\textbf{0.9456}&0.3674&0.7486\\
    \midrule
    $v_{dec} + v_{role} + v_{evid}$ & \textbf{0.7102} & 0.9342 & \textbf{0.3712} & 0.7641\\
    \bottomrule
  \end{tabular}
    \caption{Ablation of steering directions on a TREC DL 2019 validation fold with Llama-3.1-8B-Instruct. We report nDCG and MRR at a cutoff of 10, xianMAP at 100.}
  \label{tab:ablation_study}
  \vspace{-2mm}
\end{table}

%% file: table/geometry_control.tex
\begin{table}[t]
    \centering
    \small
    \setlength{\tabcolsep}{3pt}
    \begin{tabular}{lccc}
    \toprule
    Method & $\Delta$nDCG@10 & Disp. $\downarrow$ & Var. $\downarrow$\\
    \midrule
    RankSteer & \textbf{+0.0134} & \textbf{-63\%} & -10.6\%\\
    Random Direction & -0.0009 & -1\% & -0.2\%\\
    Variance-matched Comp. & -0.0020 & -2\% & \textbf{-12.0\%}\\
    \bottomrule
    \end{tabular}
    \caption{Control experiments testing whether geometric improvements arise from learned steering directions rather than generic compression. Results are averaged over 38 queries on TREC DL-2019 with Llama-3.1-8B.}
    \label{tab:geometry_control}
\end{table}

%% file: table/fine_tuned.tex
\begin{table}[t]
  \centering
  \footnotesize
  \setlength{\tabcolsep}{4pt}
  \begin{tabular}{c|ccc|c}
    \toprule
    Model & nDCG@10 & MRR@10 & MAP & BA\\
    \midrule
     Vanilla & \underline{0.6715} & 0.9079 & \underline{0.3807} &0.7694\\
     Fine-tuned & 0.6773 & \textbf{0.9430} &0.3783& 0.7650 \\
     Steered &\textbf{0.7040}&\textbf{0.9430}&\textbf{0.3831}&0.7513\\
    \bottomrule
  \end{tabular}
  \caption{Comparison of steering and fine-tuned model on the same anchor set using Llama-3.1-8B-Instruct.}
  \label{tab:fine_tuned}
\end{table}

%% file: table/collinearity.tex
\begin{table}[t]
\centering
\small
\begin{tabular}{llcc}
\toprule
\textbf{Backbone} & \textbf{Base} & \textbf{mean $|\cos|$} & \textbf{residual retained} \\
\midrule
Llama-3.1-8B & Ref  & 0.099 & 0.988 \\
Llama-3.1-8B & Role & 0.026 & 0.999 \\
Qwen2.5-7B   & Ref  & 0.074 & 0.989 \\
Qwen2.5-7B   & Role & 0.039 & 0.998 \\
Mistral-7B   & Ref  & 0.121 & 0.981 \\
Mistral-7B   & Role & 0.091 & 0.991 \\
\bottomrule
\end{tabular}
\caption{Collinearity between $v_{\mathrm{rel}}$ and $v_{\mathrm{dec}}$, averaged over layers and cross-validation anchor folds.}
\label{tab:collinearity}
\end{table}

%% file: table/role_tokens.tex
\begin{table}[t]
\centering
\small
\setlength{\tabcolsep}{4pt}
\renewcommand{\arraystretch}{1.15}
\begin{tabular}{@{}l l@{}}
\toprule
\textbf{Positive} & \textbf{Negative} \\
\midrule
\multicolumn{2}{l}{\textbf{Llama-3.1-8B-Instruct}} \\
\midrule
(brilliant, can, clearly)   & (awful, will, terribly) \\
(logical, can, carefully)   & (incorrect, will, falsely) \\
(superb, can, swiftly)      & (unreliable, will, wrongly) \\
\midrule
\multicolumn{2}{l}{\textbf{Qwen2.5-7B-Instruct}} \\
\midrule
(reliable, can, carefully)  & (messy, will, mistakenly) \\
(expert, can, swiftly)      & (problematic, will, poorly) \\
(capable, can, clearly)     & (awful, will, incorrectly) \\
\midrule
\multicolumn{2}{l}{\textbf{Mistral-7B-Instruct-v0.3}} \\
\midrule
(superb, can, clearly)      & (flawed, will, incorrectly) \\
(expert, can, perfectly)    & (messy, will, improperly) \\
(capable, can, carefully)   & (clumsy, will, sadly) \\
\bottomrule
\end{tabular}
\caption{Model-specific role token triplets (adjective, modal, adverb) used to construct contrastive role-play prompts, following \citet{wang2025role}. Each model uses three positive and three negative triplets.}
\label{tab:role_tokens}
\end{table}

%% file: table/compare_other.tex
\begin{table*}[ht]
  \centering
  \footnotesize
  \setlength{\tabcolsep}{4pt}
  \begin{tabular}{c|c|c|cccccccc|c}
    \toprule
     & & \textbf{TREC DL} & \multicolumn{9}{c}{\textbf{BEIR}} \\
    \midrule
     & \textbf{Methods} & \textbf{dl20} & \textbf{covid} & \textbf{touche} & \textbf{signal} & \textbf{news} & \textbf{scifact} & \textbf{robust04} & \textbf{dbpedia} & \textbf{nfcorpus} & \textbf{Avg} \\
    \midrule
     & BM25 & 0.4796 & 0.5947 & 0.4422 & 0.3304 & 0.3952 & 0.6789 &0.4070 & 0.3180 & 0.3218 & 0.4360\\
    \midrule
    & pairwise & 0.6098 & 0.7835& \underline{0.2578}& 0.2891 & \underline{0.4700} & \textbf{0.7032}$^\dagger$ & 0.5115 & \underline{0.3800}$^\dagger$ & 0.3658& \underline{0.4701} \\
    & setwise & 0.6112 &0.7789 & 0.2470 & \underline{0.2946} & 0.4614 & \underline{0.6808} &0.4965$^\dagger$ & 0.3774$^\dagger$ & 0.3494 & 0.4608 \\
    & listwise & \textbf{0.6587}$^\dagger$ & \textbf{0.8112} & \textbf{0.2744} & \textbf{0.3107} & \textbf{0.4833} & 0.6762 & \underline{0.5201} & \textbf{0.4189}$^\dagger$ & \underline{0.3690} & \textbf{0.4830}\\
    \cmidrule(lr){2-2}\cmidrule(lr){3-3}\cmidrule(lr){4-12}
    \rowcolor{gray!12} \cellcolor{white} & RankSteer-Ref & 0.5936 & \underline{0.7942} & 0.2443 & 0.2893 & 0.4660 & 0.6561 & 0.5126 & 0.3598 &0.3575 & 0.4600 \\
    \rowcolor{blue!8} \cellcolor{white} \multirow{-5}{*}{\rotatebox{90}{Llama-3.1-8B}} & RankSteer-Role & \underline{0.6461} & 0.7863 & 0.2445 & 0.2779 & 0.4462 & 0.6498 & \textbf{0.5207} & 0.3752 & \textbf{0.3732} & 0.4588\\
    \midrule
    & pairwise & 0.5559$^\dagger$& 0.7966& \underline{0.3099}$^\dagger$ & \underline{0.3036} & \underline{0.4854} &\textbf{0.6655}$^\dagger$ & 0.4981 & 0.3354$^\dagger$ & 0.3541&\underline{0.4696} \\
    & setwise &0.6171 & \textbf{0.8160} & 0.2879$^\dagger$ & 0.2919 & 0.4520 & 0.6260$^\dagger$ & 0.4813$^\dagger$ & 0.3953 & 0.3615& 0.4640\\
    & listwise & \textbf{0.6491} & 0.8006 & \textbf{0.3521}$^\dagger$ & \textbf{0.3268}$^\dagger$ & \textbf{0.5092}$^\dagger$ & \underline{0.6584}$^\dagger$ & \textbf{0.5359}$^\dagger$ & \textbf{0.4425}$^\dagger$ & \textbf{0.3713}$^\dagger$ &\textbf{0.4996}\\
    \cmidrule(lr){2-2}\cmidrule(lr){3-3}\cmidrule(lr){4-12}
    \rowcolor{gray!12} \cellcolor{white} & RankSteer-Ref & 0.5407 & 0.7729 & 0.2965 & 0.2849 & 0.4432 & 0.6127 & 0.4892 & 0.3375 & 0.3422 & 0.4474\\
    \rowcolor{blue!8} \cellcolor{white}  \multirow{-5}{*}{\rotatebox{90}{Mistral-7B}} & RankSteer-Role & \underline{0.6355} & \underline{0.8019} & 0.2254 & 0.2877 & 0.4441 & 0.5664 & \underline{0.5123} & \underline{0.3965} & \underline{0.3623} & 0.4496 \\
    \midrule
     & pairwise & 0.6590& 0.7989 & \underline{0.3038}$^\dagger$ & \underline{0.3111} &\underline{0.4968} & \underline{0.6968}$^\dagger$& \underline{0.5415}$^\dagger$ & 0.4134 & \underline{0.3796} & \underline{0.4927} \\
    & setwise & 0.6531 & \underline{0.8099}& 0.2883& 0.2999 & 0.4788 & 0.6800 & 0.5210 & 0.4074 & 0.3786 & 0.4830\\
    & listwise & \textbf{0.6785} & \textbf{0.8234}$^\dagger$ & \textbf{0.3184}$^\dagger$ & \textbf{0.3363}$^\dagger$ & \textbf{0.4989} & \textbf{0.7107}$^\dagger$ & \textbf{0.5438}$^\dagger$ & \textbf{0.4324}$^\dagger$ & \textbf{0.3905}$^\dagger$ &\textbf{0.5068} \\
    \cmidrule(lr){2-2}\cmidrule(lr){3-3}\cmidrule(lr){4-12}
    \rowcolor{gray!12} \cellcolor{white}  & RankSteer-Ref & 0.6475 & 0.7925 & 0.2746 & 0.3017 & 0.4913 & 0.6778 & 0.5226 & \underline{0.4148} & 0.3759 & 0.4814 \\
    \rowcolor{blue!8} \cellcolor{white} \multirow{-5}{*}{\rotatebox{90}{Qwen2.5-7B}} & RankSteer-Role & \underline{0.6594} & 0.7678 & 0.2356 & 0.2793 & 0.3807 & 0.4133 & 0.4338 & 0.3749 & 0.3697 & 0.4069\\
    \bottomrule
  \end{tabular}
  \caption{Comparison of our RankSteer method to other ranking methods, in terms of nDCG@10. The best result per dataset and LLM is marked with \textbf{boldface}; \underline{underline} indicates the second-best result. $\dagger$ denotes a significant difference from the corresponding RankSteer variant (two-tailed paired $t$-test, $p<0.05$; Ref for Llama/Qwen, Role for Mistral). Direction is omitted since nDCG@10 values indicate which method is higher.}
    \label{tab:othercomparisons}
\end{table*}